\documentclass[aps,floats,twocolumn,preprintnumbers,eqsecnum,nofootinbib]{revtex4-1}

\usepackage{amsmath}
\usepackage{amssymb}
\usepackage{amsthm}
\usepackage{float}
\usepackage{graphicx}
\usepackage{subfigure}

\newcommand{\bea}{\begin{eqnarray}}
\newcommand{\eea}{\end{eqnarray}}
\newcommand{\beq}{\begin{equation}}
\newcommand{\eeq}{\end{equation}}
\newcommand{\lapse}{\alpha}

\begin{document}

\title{Quantum Back Reaction to asymptotically AdS Black Holes}
\author{Kazumi Kashiyama}
\email{kashiyama@tap.scphys.kyoto-u.ac.jp}
\author{Norihiro Tanahashi}
\email{tanahashi@tap.scphys.kyoto-u.ac.jp}
\affiliation{Department of Physics, Kyoto University, Kyoto 606-8502, Japan}

\author{Antonino Flachi}
\email{flachi@yukawa.kyoto-u.ac.jp}
\author{Takahiro Tanaka}
\email{tanaka@yukawa.kyoto-u.ac.jp} 
\affiliation{Yukawa Institute for Theoretical Physics, Kyoto University, Kyoto 606-8502, Japan}

\preprint{KUNS-2237,~YITP-09-63}

\begin{abstract}
We analyze the effects of the back reaction due to a conformal field theory
 (CFT) on a black hole spacetime with negative cosmological constant.
We study the geometry numerically obtained by taking into account 
the energy momentum tensor of CFT approximated by a radiation fluid.
We find a sequence of configurations without a horizon in
 thermal equilibrium ({\it CFT 
 stars}), followed by a sequence of configurations with a horizon. 
We discuss the thermodynamic properties of the system and how
 back reaction effects alter the space-time structure.
We also provide an interpretation of the above sequence of solutions in terms
 of the AdS/CFT correspondence. The dual five-dimensional description is
 given by the Karch-Randall model, 
in which 
 a sequence of five-dimensional floating
 black holes followed by a sequence of brane
 localized black holes correspond to the above solutions.  
\end{abstract}
\maketitle
\vspace{2mm}

\section{Brane world black holes and the AdS/CFT correspondence}\label{intro}

\renewcommand{\thefootnote}{}

The AdS/CFT conjecture relates the gravitational dynamics of a
$(d+1)$-dimensional AdS spacetime to a $d$-dimensional conformal field
theory (CFT), and it was initially formulated as a correspondence
between type IIB supergravity on $\mbox{AdS}_5\times \mbox{S}^5$ and
${\cal N}=4$ $U(N)$ super Yang-Mills (SYM) theory, with coupling $\hat{g}$
and 't Hooft parameter $\lambda = \hat{g}^2N$, related to the
supergravity parameters by
\footnote{
\\
\hspace{-10pt}
${}^1$
This formula is different from the ordinary AdS/CFT dictionary, 
${\ell^3/G_5}={2N^2/\pi}$, since we focus on models with two AdS bulk regions 
and hence the degrees of freedom of the CFT, $N^2$, is doubled from $\pi \ell^3/2G_5$ to $\pi \ell^3/G_5$.
}
\beq
\ell=\lambda^{1/4}\ell_s~,~~~~{\ell^3\over G_5}={N^2\over
\pi}~.\label{N}
\eeq
In the above formulas $\ell_s$ is the string length, $\ell$ and $G_5$
are the five-dimensional AdS curvature length and Newton constant,
respectively~\cite{Mal,Wit,Gub}.
The correspondence relates the supergravity partition function in
AdS$_5$ to the generating functional $W_{CFT}$ of connected Green's functions for the CFT on the boundary, and it has a very interesting connection with the Randall-Sundrum (RS) model~\cite{RS}. 

\renewcommand{\thefootnote}{\arabic{footnote}}

The RS model consists of two copies of a part of AdS$_5$-like
 spacetime. The boundaries of the copies are glued with a 
positive tension brane. 
The model is described by the following action:
\beq
S_{RS}=S_{EH}+S_\text{brane}+S_M~,
\label{RSaction}
\eeq
where $S_{EH}$ is the five-dimensional Einstein-Hilbert action, 
$S_\text{brane}$ represents the action of the brane with tension $\sigma
= 3/4\pi G_5\ell$, and $S_{M}$ describes matter confined on the brane. 
On the four-dimensional brane, asymptotically flat spacetime is realized
 by tuning the brane tension relative to the negative bulk cosmological
 constant. 
The Standard Model is localized on the brane, and the
observed four-dimensional nature of gravity arises owing to the
presence of a localized graviton zero mode~\cite{GT}. 

References~\cite{Gubser2,Hawking} provided an interpretation of the RS model
in terms of the AdS/CFT correspondence. The correspondence implies 
\beq
S_\text{RS} = -{\ell\over 16 \pi G_5}\int d^4x \sqrt{g} R + 2W_{CFT}+S_M~,
\eeq
indicating that the classical gravity in the RS model is dual to
four-dimensional gravity coupled to a cutoff CFT. 
From the above action, the effective four-dimensional Newton constant 
is read as 
\begin{equation}
 G_4={G_5\over \ell}.
\end{equation}
In the linearized, weak gravity regime various results clearly support 
this conjecture~\cite{Hawking,Duff,TanakaFRWGW,PujolasDW}. 
On the other hand, although some evidence for the conjecture exists 
({\it e.g.}~\cite{tanaka}), 
things become more complicated when trying to extend the correspondence
to the non-linear regime in general~\cite{Shiromizu,GarrigaSasaki}, 
and to black holes in particular. In this paper we will be concerned
with the latter case. 

Let us summarize the present understanding of black hole solutions in
the RS model. 
In this model no {\it large}, stable, static black hole solution
localized on the brane has so far 
been found, whereas {\it small} localized solutions have been
constructed numerically \cite{Kudo1} 
for black holes with size smaller than the curvature scale $\ell$
\footnote{It is fair to mention that the existence of such solutions is
still controversial~\cite{Yoshino}.}. 

This situation can be interpreted in the light of the AdS/CFT correspondence, and in 
Refs.~\cite{emparan,tanaka} it has been conjectured that large stable
black holes localized on the brane 
{\it do not exist} in the RS model. 
The intuitive picture is as follows.
Consider a four-dimensional black hole with CFT. This black hole
will evaporate into CFT modes.
If the correspondence is valid also in this situation, this evaporation 
process must be equivalent
to a classical five-dimensional dynamical phenomena.
This may imply that there is no stationary black hole solution in the
five-dimensional RS model,
and that the five-dimensional black hole ``evaporates'' by a classical
process.
Note that existence of the numerical solutions of Ref.~\cite{Kudo1},
describing small black holes, is not in 
contradiction with the above statement since the correspondence is not
expected to hold below the cutoff length scale of the CFT, which is
of order of the AdS curvature scale $\ell$.

Black holes {\it floating} in the bulk 
are also expected to exist~\cite{tanaka}, although no solution of this sort has been
found. 
Such floating black holes also cannot be large 
for the following reason. 
In the RS model,
the gravitational force between the brane and a particle in the bulk is repulsive. 
Writing the metric in Poincar\'{e} coordinates,
\beq
ds^2 = dy^2 +e^{-2y/\ell}\left(-d\bar t^2+d{\bf \bar x}^2\right)~,
\eeq  
one can see that the acceleration of a particle is 
$a=-\partial_y \log\sqrt{-g_{\bar t\bar t}}=1/\ell$ and independent of
$y$. 
The only force that compensates such repulsive force is the self-gravity of
the mirror image of the particle 
on the other side of the brane. From the above observation, we expect
that, as $\ell$ decreases, 
the equilibrium position of the floating black hole should move towards
the brane. However, the 
attractive force between the black holes is at most of $O(1/r_{\rm h})$, with
$r_{\rm h}$ being the horizon size. 
If $r_{\rm h}\gg \ell$, such attractive force will not be sufficient to
cancel the repulsive force from the brane. 
Thus, large black holes will necessarily touch the brane. 

Although the difficulty in constructing large
localized (or floating) black hole
solutions in the RS model synchronizes with the prediction from 
the AdS/CFT correspondence, it is also true that larger black holes 
become more difficult to construct simply for a technical reason 
because two different scales, the bulk curvature scale 
$\ell$ and the black hole size, should be resolved simultaneously. 
Therefore it is difficult to prove the absence of solutions
numerically. 
Then, one of the authors proposed 
to study black holes in Karch-Randall (KR) model~\cite{KR}, in which 
the brane tension is chosen to be less than the fine-tuned value of 
the RS model~\cite{tanaka2}. 
Unperturbed background bulk geometry is AdS, 
which is conveniently described by 
\beq
ds^2= dy^2+\ell^2 \cosh^2\left(y/\ell\right) ds^2_{AdS_4}~,
\label{KRcoord}
\eeq
where $ds^2_{AdS_4}$ 
is the line element of four-dimensional AdS
spacetime with unit curvature:
\begin{equation}
ds^2_{AdS_4}=
 -\left(1+{\bar r^2}\right)d\bar t^2+\left(1+{\bar r^2}\right)^{-1}d\bar
 r^2+\bar r^2 d\Omega_2^2.
\end{equation}
Contrary to the Poincare chart convenient for RS model, 
in this chart the warp factor is not monotonic 
but has a minimum at $y=0$. The position of the
brane is specified by $y=y_b$, and $y_b$ is determined by the condition 
\beq
\sigma = -{3\over 4\pi G_5  \ell} \tanh\left(y_b/\ell\right)~.  
\eeq
The RS limit is obtained by letting $y_b \rightarrow -\infty$. 

\begin{figure}[h]
    \begin{center}
      \resizebox{80mm}{!}{\includegraphics[angle=270]{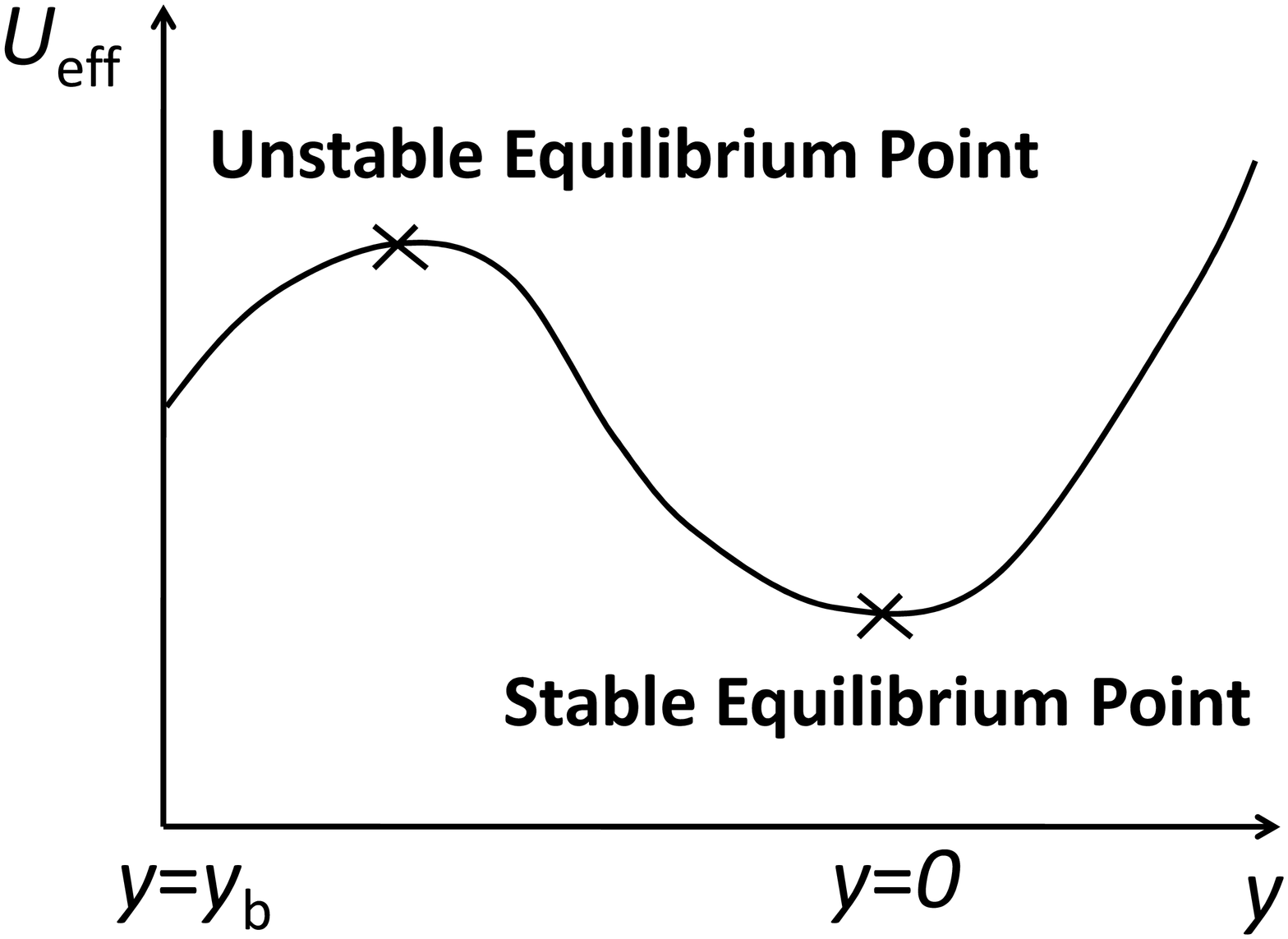}} \\
  \caption{Effective potential in the KR model.}\label{effpot}
    \end{center}
\vspace{4mm}
\end{figure}

Let us consider a small black hole floating in the bulk of the KR
model. Following the same analogy as before, we can look at the
acceleration of a small mass particle. In this case a particle feels a
potential
\beq
U_\text{eff} = \log \left(\cosh {y\over \ell}\right)+U_\text{sg}~,
\eeq
where $U_\text{sg}$ is the self-gravitational part caused by its own mirror 
image on the other side of the brane. 
The profile of $U_\text{eff}$ is illustrated in Fig.~\ref{effpot} and suggests that 
there will be two small black hole solutions: 
an unstable one close to the UV brane, and a stable one near $y=0$, far
from the UV brane. In the RS limit $\delta \sigma \rightarrow 0$, the
latter is infinitely far from the brane and hence it does not exist. It is natural to imagine 
that the stable floating black holes also touch the brane when the size becomes big enough. 

According to the AdS/CFT correspondence, we may expect that a five-dimensional
black hole in the KR model will be dual to 
some object in the four-dimensional gravity coupled to CFT with 
negative cosmological constant~\cite{tanaka2}.
Naive expectation is that
a brane-localized black hole and a floating black hole in the KR model
are, respectively, dual to a 
four-dimensional black hole with back reaction of CFT halo and a star
composed of CFT, which we refer to in this paper as 
a quantum black hole and a CFT star.  
If this duality is really the case, we can
examine black holes in the KR model by analyzing the four-dimensional system.

On the four-dimensional brane in the KR model, asymptotically AdS
spacetime is realized. Under the restriction to 
static and spherically symmetric configurations, the four-dimensional 
metric is in general written as 
\begin{eqnarray}
ds^2=-\alpha(r)^2 dt^2+V(r)^{-1}dr^2+r^2d\Omega_2^2~.
\label{4dmetric}
\end{eqnarray}
As we consider an equilibrium configuration 
with a black object in the bulk, 
the corresponding CFT is also expected to be in thermal 
equilibrium at a finite temperature. 
The local temperature of CFT in equilibrium red-shifts as 
\begin{equation}
 T_{\rm local}(r)=T/\lapse(r),  
\end{equation} 
where the global temperature of the system $T$ is defined with respect to 
the time-like Killing vector $\partial/\partial t$. 

Let us consider a quantum black hole in the above thermal AdS spacetime.
For black hole configurations in equilibrium 
the appropriate vacuum state will be the Hartle-Hawking state. 
In the asymptotically flat case, back reaction due to CFT is too strong to 
keep the asymptotic structure of the spacetime unchanged (the total mass diverges). 
In the asymptotically AdS case, a 
non-zero cosmological constant changes the situation dramatically.
Since the lapse function in AdS behaves as $\lapse\sim r/L$
for large circumferential radius $r$, where $L$ is 
the four-dimensional AdS curvature scale, the temperature and 
hence the energy density of thermal CFT decrease rapidly for $r\gg L$. 
This reduces the effects of the back reaction. 
If the black hole size is large, 
the energy density due to CFT 
will stay negligibly small at any radius. 
Whilst, if the size of the black hole is small,
the back reaction becomes important and a static black hole solution 
becomes non-trivial. 
Roughly speaking, such a small black hole will be unstable against the
CFT back reaction and will `evaporate' into a CFT star of the same mass.

The sequence of the CFT stars can be tagged by the central density, 
and the end-point of the sequence corresponds to a star with singular
central density and the lapse vanishing at the center.
Thus, this sequence of the CFT stars will naturally flow into the
sequence of quantum 
black holes, whose starting-point corresponds to a small black hole in
the limit of zero horizon radius.

We can interpret the sequence of four-dimensional quantum black holes and CFT 
stars from a five-dimensional view point as follows.
At the transition point of the sequence, 
the lapse vanishes at the center of the system. 
This four-dimensional configuration corresponds to a five-dimensional
black hole floating in the 
bulk and just touching the brane, since the lapse vanishes at the touching point for this 
five-dimensional configuration too.
In this way, we may speculate that
the sequence of floating black holes corresponds to the sequence of CFT
stars, while the sequence of brane-localized black holes corresponds to
the sequence of quantum black holes. 

In this paper, we will present our investigation concerning the
four-dimensional asymptotically AdS quantum black holes and CFT stars with the aim of clarifying
the phase diagram structure of black objects in the KR model.
We will give quantitative estimates of the characteristic quantities of
the model by explicitly 
constructing equilibrium configurations in the dual picture 
described by four-dimensional
gravity with CFT correction.
In Sec.~\ref{sec:CFT}, we will show that the effects of CFT can be
properly approximated by a radiation fluid.
We analyze properties of CFT in Schwartzschild AdS spacetime and
give the conditions for the 
radiation fluid approximation to CFT to be applicable. 
We will show that those four-dimensional objects 
in equilibrium state can be well
approximated by this approximation,  
as long as we restrict our interest to 
the range of parameters where the correspondence is expected to be valid. 
In Sec.~\ref{sec:float}, we will illustrate our method to construct equilibrium configurations of
four-dimensional self-gravitating CFT and study its basic properties
that can be derived analytically.
The full numerical analysis will be given in Sec.~\ref{sec:numeric}.
Based on the above results, we will finally discuss the implications
for the KR model via the AdS/CFT correspondence in Sec.~\ref{sec:AdS/CFT}
and summarize the paper in Sec.~\ref{sec:summary}.

For notation convenience, we set $G_4$ to unity from Sec.~\ref{sec:CFT} to Sec.~\ref{sec:numeric}.

\section{CFT Energy-Momentum Tensor and Radiation fluid approximation}
\label{sec:CFT}

In order to study the effects of the back reaction, explicit knowledge
of the quantum energy-momentum tensor of the CFT in the Hartle-Hawking
vacuum state is necessary. The computation is, however, technically very
complicated and, apart from Ref.~\cite{flachitanaka} where the vacuum
polarization has been obtained for a conformal scalar field, we are not aware of other relevant 
results for Schwarzschild AdS black holes in the literature. Even had we obtained the exact
expression for the energy-momentum tensor, additional problems would
arise in solving the Einstein equations self-consistently. 
The energy momentum tensor of CFT effectively contains higher 
derivatives of the metric functions, and 
those terms will introduce spurious solutions and make the choice of
boundary conditions quite non-trivial. 
For this reason, as a first step, it seems natural to look for
a simplified scheme to take into account the quantum back reaction 
approximately. 
In this paper we propose to use the radiation fluid approximation, 
which makes it easy to study back reaction effects of CFT on the
spacetime structure. 
In this section we evaluate the CFT energy-momentum tensor using 
Page's approximation~\cite{page} on the Schwarzschild AdS black hole background~\cite{gregory}, 
and compare the results with those obtained by the radiation fluid 
approximation. 

Before introducing Page's approximation, it is instructive to discuss
the relevant length scales. 
Since the CFT is scale invariant, the
only scales that characterize the system are (i) geometrical length
scales of the space-time, such as the distance from the BH horizon
radius $r-\bar{r}_{\rm h}$ or the curvature scale, 
and (ii) the scale related to the local temperature of the system
$1/T_{\rm local}$. 
For high enough temperatures, $1/T_{\rm local}$ becomes the only relevant
length scale of the system, except for the vicinity of the horizon.  
In this case, from the symmetry, the energy momentum tensor of CFT
should follow a Stephan-Boltzmann law, 
\begin{equation}\label{RF}
T^{\mu}{}_{\nu}=\frac{\pi^2}{30}g_{\rm eff} 
 T_{\rm local}^4(\delta^{\mu}{}_{\nu}-4\delta^{\mu}{}_{0}\delta^{0}{}_{\nu})~,
\end{equation}
where $g_{\rm eff}$ represents the effective number of degrees of freedoms. 
Expression (\ref{RF}) claims 
that a thermal CFT can be approximated by a radiation fluid 
when the red-shifted temperature of the system is high enough. 

The procedure is, however, not straightforward, since the radiation
fluid approximation breaks down near the horizon, due to the fact that
the local temperature $\propto\lapse^{-1}$ diverges there. 
In order to remove this pathology, we
need to consider the quantum contribution to the energy momentum tensor,
and will use Page's approximation for this purpose. 
We split the genuine energy density into two parts, $\rho=\rho_{\rm r}+\rho_{\rm q}$, 
where $\rho_{\rm r}$ corresponds to 
a classical radiation fluid contribution and $\rho_{\rm q}$ to a quantum
contribution that is defined by the remainder of this section. 

As an example, we consider a conformal scalar field on Schwarzschild AdS 
background. In this case Page's approximation is known to be 
equivalent to the fourth order WKB approximation~\cite{anderson}. 
In order to take into account all the degrees of freedom of SYM, 
conformal spinor and vector contributions should be included. However,
Page's approximation produces unphysical divergences on the
event horizon for these cases, and to include such contributions
rigorously, a more sophisticated numerical method is
needed~\cite{anderson}. 
We do not pursue this rather technically complicated issue here. 
For a conformal scalar field, 
the classical part $\rho_{\rm r}$ described by a radiation fluid is 
given by
\bea
\rho_{\rm r}
={r^2  \over K \bar{r}_{\rm h}^4}(1+3\bar{r}_{\rm h}^2/L^2)^4~,
\label{explr}
\eea
while the quantum part $\rho_{\rm q}$ is computed by using Page's
approximation as 
\bea
\rho_{\rm q}
=-{1\over K} \sum_{i=0}^{9} a_i r^{i-6}~,
\label{explrho}
\eea
with
\bea
K\equiv 
7680 \pi^2\left(r-\bar{r}_{\rm h}\right)^{2}\left\{1+\left(r^2+\bar{r}_{\rm h} r+\bar{r}_{\rm h}^2\right)/L^2\right\}^{2}
\nonumber~.
\eea
The coefficients $a_i$ are polynomials in $\chi\equiv \bar{r}_{\rm h}/L$ of degree smaller than 9:
\begin{align}
a_0 &= 33 \bar{r}_{\rm h}^4
\left(1+\chi^2\right)^4
~,
&
a_1 &= -72 \bar{r}_{\rm h}^3
\left(1+\chi^2\right)^3
~,\nonumber\\
a_2 &= 40 \bar{r}_{\rm h}^2
(1+\chi^2)^2
~,
&
a_3 &= -88 \bar{r}_{\rm h} 
\chi^2\left(1+\chi^2\right)^3
~,\nonumber\\
a_4 &= 104
\chi^2\left(1+\chi^2\right)^2
~,&
a_5 &= 0~, \nonumber\\
a_6 &= {104\over L^2}
\chi^2\left(1+\chi^2\right)
~,&
a_7 &= -{32\over L^3}
\chi\left(1+\chi^2\right)
~,\nonumber\\
a_8 &= {16\over L^4}~,&
a_9 &= {32\over L^5}
\chi\left(1+\chi^2\right)
~.\nonumber
\end{align}

\begin{figure}[h]
    \begin{center}
      \resizebox{80mm}{!}{\includegraphics{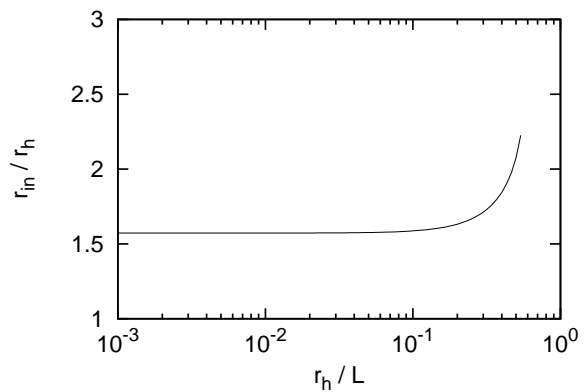}} \\
   \caption{Relation between the horizon
     radius $\bar{r}_{\rm h}$ and critical radius $r_{\rm in}$. We have set $c=1/2$ and $l/L=10^{-8}$.}
\label{setuzoku}
    \end{center}
\vspace{4mm}
\end{figure}

Let us consider the quantum effects in the near horizon region first.
We define an inner critical radius $r_{\rm in}$  
as the radius at which the equality
\begin{equation}\label{hantei}
|\rho_{\rm q}| = c\rho_{\rm r}
\end{equation} 
is first satisfied, with $c$ being a constant of order $O(1)$. 
For $r<r_{\rm in}$ the quantum contribution 
dominates the classical radiation part. 
The result is not so sensitive to the choice of $c$ as long as it is $O(1)$. 
Figure \ref{setuzoku}
illustrates the relation between the horizon radius $\bar{r}_{\rm h}$ and the inner
critical radius $r_{\rm in}$:
when $\bar{r}_{\rm h}$ is small compared with the four-dimensional curvature scale
$L$, $r_{\rm in}/\bar{r}_{\rm h}$ is 
approximately constant of $O(1)$ ({\it e.g.}~$r_{\rm in} \sim 3\bar{r}_{\rm h}/2$ for $c=1/2$); 
when we increase $\bar{r}_{\rm h}$ beyond $L$, $r_{\rm in}/\bar{r}_{\rm h}$ begins to
increase and finally
solutions of Eq.~(\ref{hantei}) cease to exist. 
No inner critical radius can be defined for larger 
$\bar{r}_{\rm h}$.

\begin{figure}[h]
    \begin{center}
      \resizebox{80mm}{!}{\includegraphics{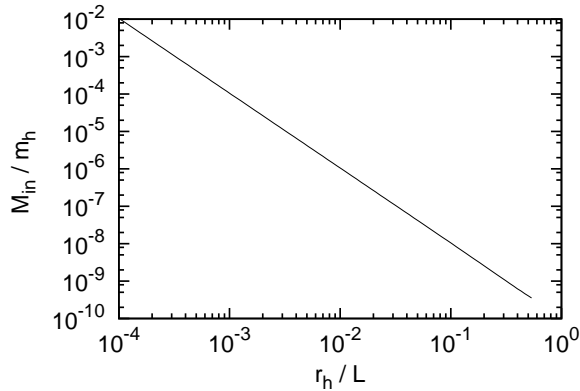}} \\
   \caption{Ratio of the total energy of the CFT inside the critical radius and the black hole horizon radius. We set $\ell^2/L^2=10^{-8}$.}
\label{inside}
    \end{center}
\vspace{4mm}
\end{figure}

Having fixed the critical radius as above, we can discuss the strength
of the reaction due to the CFT in the region $r<r_{\rm in}$ by
comparing the total energy of
the CFT within the critical radius $M_{\rm in}$ with the black hole mass
$m_{\rm h}$. When the size of the black hole is small, $M_{\rm in}$ is estimated
as $M_{\rm in}\sim (\pi^2 /30) g_{\rm eff} T_{\rm local}^4 \bar{r}_{\rm h}^3\sim \ell^2/\bar{r}_{\rm h}$, where 
we have substituted 
\begin{eqnarray}\label{geff}
 g_{\rm eff}=\frac{3}{4}\cdot15N^2~,
\end{eqnarray} 
which is the value for ${\cal N}=4$ $U(N)$ SYM theory, and used the relation (\ref{N}).
The factor $3/4$ in Eq.~(\ref{geff}) is the empirical factor that explains the discrepancy 
between results for CFTs in strong and weak coupling cases~\cite{3/4factor}.  
Hence, we have 
\beq
{M_{\rm in}\over m_{\rm h}} \sim \left(\ell\over \bar{r}_{\rm h}\right)^2~.
\label{compareM}
\eeq
It is easy to see that, as long as $\ell / \bar{r}_{\rm h}$ is small, the above
ratio is also small, and the contribution to the total mass of 
CFT living inside the inner critical radius is negligible. The
cases with $\ell > \bar{r}_{\rm h}$ are beyond the range of applicability of the AdS/CFT correspondence 
and are outside the parameter region of our interest. 
For illustration, Fig.~\ref{inside} shows the ratio
$M_{\rm in}/m_{\rm h}$ with respect to the horizon radius. 

Let us move on to the quantum effect in the asymptotic region next.
At a large distance, 
the classical radiation part of the energy density $\rho_{\rm r}$ behaves as
$1/r^4$, whilst the quantum part $\rho_{\rm q}$ 
behaves as $1/r^3$, as is seen from Eq.~(\ref{explrho}). 
Hence, the quantum part dominates above an outer critical 
radius, $r_{\rm out}$, which is defined as before by Eq.~(\ref{hantei}). 
By comparing Eqs.~(\ref{explr}) and (\ref{explrho}), we find 
$r_{\rm out}\sim L \left(L / \bar{r}_{\rm h}\right)^{5}$ for small black
holes with $\bar{r}_{\rm h} \ll L$. 
This quantum part $\rho_{\rm q}$ in the outer region 
gives a non-negligible contribution for large $r$. 
In fact, by taking $\rho_{\rm q}$ into account, the mass measured at $r$, 
$M(r)$, will be modified as follows.
Eq.~(\ref{explrho}) suggests that the leading term of $\rho_q$ behaves as 
$\sim -M(r)\ell^2/L^2 r^3$ when we consider the back reaction of the CFT to the 
background geometry%
\footnote{Note that Eq.~(\ref{explrho}), which gives $\rho_{\rm q}\sim-m_{\rm h}/L^2r^3$,
is for one conformal scalar while we consider $N^2\sim \ell^2$ degrees of freedom here.
Adding to that, the back reaction of the CFT to the background geometry will change its behavior from 
$-m_{\rm h}\ell^2/L^2r^3$ to $-M(r)\ell^2/L^2r^3$.
}.
Then, $M(r)$ will be modified as
\begin{equation} 
M(r)\sim \int^{r} r'^{\,2}\rho_{\rm q}dr'
\sim M(r_{\rm out})\times \left(\frac{r}{r_{\rm out}}\right)^{-\ell^2/L^2},
\label{Mvariance}
\end{equation}
where $M(r_{\rm out})$ is the mass in the region $r<r_{\rm out}$.
Hence, the effect of $\rho_{\rm q}$ significantly alter
the total mass $M$ from the value for the bare black hole at a very large
distance. 
The above result can be interpreted as the mass screening 
effect due to the non-zero graviton mass of
$O(\ell/L^2)$~\cite{porrati,AdSDuff}
\footnote{
When the graviton has small non-zero mass $m_g$,
roughly speaking, the metric perturbation $h_{\mu\nu}$ obeys~\cite{KR}
\begin{equation}
 \left(r^{-2}\partial_r r^4 \partial_r -m_g^2L^2  \right)(h_{\mu\nu}/r^2)=0~,
\notag
\end{equation}
which implies $h_{\mu\nu}\propto r^{-1-(1/3)(m_g L)^2}$ and hence $M\propto r^{-(1/3)(m_g L)^2}$.
Hence, the leading correction due to the graviton mass of $O(\ell/L^2)$~\cite{porrati,AdSDuff},
reproduces the $r$-dependence presented in Eq.~(\ref{Mvariance}).
}. 
To avoid ambiguity of $M$ with respect to $r$, strictly speaking, we 
need to truncate the model at a 
finite radius well outside the AdS curvature
radius but before this screening effect becomes significant. 
This prescription will be justified later in Sec.~\ref{sec:AdS/CFT}.
As long as this truncated
model is concerned, we can neglect the quantum part of the energy
density $\rho_{\rm q}$ also in the region $r>r_{\rm out}$. 
Once we neglect the quantum part, this screening effect is also absent. 
Hence, in the actual computation discussed in the succeeding sections, 
where we use the radiation fluid approximation, we do not have to 
care about this truncation.  

As the size of a black hole becomes large compared with $L$, 
the interval between $r_{\rm in}$ and $r_{\rm out}$ shrinks and eventually 
disappears. Beyond that point, the classical radiation part of 
the energy density $\rho_{\rm r}$ does not dominate the quantum part for any
$r$. Hence, one may think that 
the approximating the energy momentum tensor by a radiation
fluid is not a good approximation at all.  
However, in this case the temperature is so low 
that the back reaction to the mass due to CFT is negligibly
small, as long 
as we adopt the above prescription of truncating the model at a 
finite radius before the screening effect becomes significant. 
Hence, we conclude that in all cases that we are interested in 
the radiation fluid approximation is expected to give a good
approximation to the energy momentum tensor of CFT except for the 
vicinity of the event horizon. 

\section{Boundary theory description of floating black holes}
\label{sec:float}

\subsection{CFT stars}

First, we re-examine spherically symmetric, equilibrium configuration 
of a radiation star in asymptotically AdS spacetime, which was analyzed in Ref.~\cite{PP}. 
The only difference from the literature is that the effective number of degrees of
freedom $g_{\rm eff}$ is set to a large number $(45/4)N^2$ in connection to the AdS/CFT correspondence.  

To deal with the above problem, 
it is convenient to write the metric (\ref{4dmetric}) as
\begin{equation}
ds^2=-e^{2\psi}Vdt^2+V^{-1}dr^2+r^2(d\theta^2+\sin^2\theta d\phi^2)~,
\label{four-d}
\end{equation}
with
\begin{equation}\label{V}
V=1+\frac{r^2}{L^2}-\frac{2m(r)}{r}~.
\end{equation}
We re-parametrize the time coordinate $t$ 
so as to satisfy 
\begin{equation}\label{bcpsi}
\lim_{r\to \infty}\psi(r)=0~. 
\end{equation}
In these coordinates the total mass of the system is given by 
\begin{equation}
M\equiv \lim_{r\to \infty}m(r)~,\nonumber
\label{bcm}
\end{equation}
Pressure and energy density can be written as
\begin{equation}
\rho=3P_r=3P_{\theta}=\frac{\pi^2}{30}g_{\rm eff}T_{\rm local}^4=\frac{3\pi^3
 \ell^2}{8} T^4e^{-4\psi}V^{-2}~.
\label{eos}
\end{equation} 
The Einstein equations (with $\Lambda=-3/L^2$) are
\bea
\label{Eins1}
\frac{dm}{dr}&=&4\pi r^2 \rho~,\\
\label{Einspsi}
\frac{d\psi}{dr}&=&\frac{16\pi}{3}rV^{-1}\rho~,\\
\label{Eins3}
\frac{d\rho}{dr}&=&-\frac{4\rho(m+4\pi r^3 \rho/3 +r^3/L^2)}{r^2+r^4/L^2-2rm}~.
\eea
The central density 
\begin{eqnarray}
\rho_c\equiv \rho(0),
\end{eqnarray} 
can be used to parametrize the solutions, 
and the boundary condition for $m(r)$ is specified by 
\bea
 m(0)&=&0~.
\eea
Then, integrating Eqs.~(\ref{Eins1}) and (\ref{Eins3}) from $r=0$ 
for given curvature scales $\ell$ and $L$, we obtain a one-parameter family of
non-singular equilibrium configurations labelled by $\rho_c$. 
From the boundary condition (\ref{bcpsi}) and the relation 
(\ref{eos}), we obtain the global temperature
\bea
\label{bcT}
T=\lim_{r\to \infty}\Big( \frac{8}{3\pi^3\ell^2}\rho V^2 \Big)^{1/4}~.\\ \nonumber
\eea
Finally, the solution of Eq.~(\ref{Einspsi}) is obtained algebraically
from Eq.~(\ref{eos}) as 
\begin{equation}
\psi(r)=\frac{1}{4}\ln \Big(\frac{3\pi^3 \ell^2}{8} T^4\rho^{-1} V^{-2}
 \Big), 
\end{equation}
without solving Eq.~(\ref{Einspsi}). 
Another global quantity of interest  
is the total entropy of the system, $S(L,\ell,\rho_c)$.
Once the functional dependence of $M$ and $T$ upon $\rho_c$ is
determined, $S$ can be obtained by integrating the first law of
thermodynamics,
\begin{equation}\label{first}
dS=\frac{dM}{T}
\end{equation}
for fixed $\ell$ and $L$, with $S=0$ at $\rho_c=0$. 

Notice that, 
writing the equations in terms of the rescaled variables associated 
with ``$~\tilde{}~$'', defined by 
\bea
r= L \tilde r~,\quad
\rho = L^{-2}\tilde{\rho}~,\quad
m=L\tilde m,
\eea
$L$-dependence is eliminated from the Einstein equations and also from the form of the
metric function (\ref{V}). 
Accordingly, 
the rescaled thermodynamic quantities defined by 
\begin{eqnarray}
M(L,\ell,\rho_c) &=& L \tilde{M}(L^2\rho_c)~,\label{scal1}\\
T(L,\ell,\rho_c) &=& \ell^{-1/2}L^{-1/2}\tilde{T}(L^2\rho_c)~,\label{scal2}\\
S(L,\ell,\rho_c) &=& \ell^{1/2}L^{3/2}\tilde{S}(L^2\rho_c)~,\label{scal3}
\end{eqnarray}
absorb the $\ell$-dependence present in Eq.~(\ref{eos}), too. 
Owing to the above scaling relations, we can set $L=\ell=1$ 
without loss of generality. 

Details of the numerics for the CFT star configurations will be reported
in the succeeding section along with the black hole ones. Here we wish to
close this subsection with some analytic estimates of the
thermodynamic quantities of our interest. 
As discussed in Ref.~\cite{PP}, when the radiation has negligible
self-gravity, the metric will 
be approximated by the pure AdS spacetime.
The condition for the self-gravity to be negligible can
be stated as 
\begin{equation}
  m(r)/r \ll 1, 
\end{equation}
for all $r$. 
This condition is satisfied when the central density of the radiation is
much smaller
than that corresponds to the four-dimensional AdS curvature scale, 
$(\pi^2/30)g_{\rm eff}T^4\ll L^{-2}$. 
The temperature corresponding
to the critical central density at which  
the equality $(\pi^2/30)g_{\rm eff}T^4= L^{-2}$ is satisfied  
will then be approximately given by
\begin{equation}
T \sim \ell^{-1/2} L^{-1/2}.
\end{equation}
Below this temperature, the spacetime is practically AdS, which will work as a box of
the volume of $O(L^3)$ for the radiation. Then, the total energy of the system can
be estimated easily as 
\begin{equation}
M\sim \ell^2 L^3 T^4 .
\label{Mrad}
\end{equation}
Similarly, the total entropy of the system is approximated by
\begin{equation}
S\sim \ell^2 L^3 T^3 .
\label{Srad}
\end{equation}
Substituting $T \sim \ell^{-1/2} L^{-1/2}$, 
one can estimate the 
total mass and the total entropy at the critical point where the back
reaction to the geometry becomes important as 
\bea 
M/L&\sim&1~,\\
\ell^{-1/2}L^{-3/2}S&\sim& 1~,
\eea
which is found to be consistent with the scaling relations 
(\ref{scal1}), (\ref{scal2}) and (\ref{scal3}). 
A precise evaluation of all the thermodynamical quantities will be given
later by explicit numerical computations.

\subsection{AdS Black holes with CFT back reaction}

Next, we discuss configurations with a black hole horizon. 
As we discussed in Sec.~\ref{sec:CFT}, 
fluid approximation breaks down near the horizon. 
There is a critical radius $r_{\rm in}$, and for $r<r_{\rm in}$
we cannot neglect the quantum correction to the energy momentum tensor. 
However, the role of quantum correction is simply to regularize the 
divergent energy density obtained in the fluid approximation, and 
hence it is possible to approximate solutions in this region by 
a vacuum solution of the Einstein equations, i.e. 
Schwarzschild AdS solutions. 
In the following, we will use the critical radius $r_{\rm in}$ as the
junction radius at which the Schwarzschild AdS solution for $r<r_{\rm in}$ 
is connected to the solution that includes the CFT back reaction for $r>r_{\rm in}$. 
As before, we assume that the spacetime is static and spherically
symmetric. Thus, Eqs.~(\ref{eos}), (\ref{Eins1}), 
(\ref{Einspsi}) and (\ref{Eins3}) are the same as before. 
Only the inner boundary conditions are different. 
From the continuity of the metric functions at $r=r_{\rm in}$, 
we obtain the boundary conditions,  
\bea
\label{connect1}
m(r_{\rm in})&=&m_{\rm h}~,\\
\label{connect2}
e^{\psi(r_{\rm in})}dt~&=&d\widehat{t},
\eea
where $m_{\rm h}$ is the mass parameter of the central Schwarzschild AdS metric that
describes the region $r<r_{\rm in}$, and $\hat t$ is the time coordinate 
in the inner region $r<r_{\rm in}$. 
We require that the temperature of the inner black hole solution is
equal to that of the outer thermal radiation fluid. 
Then, we obtain 
\bea
T&=&e^{\psi(r_{\rm in})} \widehat{T}~, 
\label{connect3}\\
\widehat{T}&=&\frac{L^2+3\bar{r}_{\rm h}^2}{4\pi L^2 \bar{r}_{\rm h}}~. 
\label{connect3-2}
\eea
Here $\hat T$ is the temperature defined with respect to the
timelike Killing vector $\partial/\partial \hat t$. 
The factor $e^{\psi(r_{\rm in})}$ in Eq.~(\ref{connect3}) 
takes care of the difference between
the time coordinates for $r \leq r_{\rm in}$ and for $r\geq r_{\rm in}$. 
With the above boundary conditions, Einstein's equations can be 
solved numerically and the thermodynamical quantities evaluated 
for various values of the horizon radius $\bar{r}_{\rm h}$. 
Here the scaling relations that hold in the star case 
are not fully compatible with the boundary condition (\ref{connect3}) 
with (\ref{connect3-2}), 
which requires $T$ to scale like $T=\tilde T/L$. 
Therefore we cannot completely absorb dependences on both $L$ and $\ell$ 
by the rescaling, and the dimensionless ratio $\ell/L$ remains as 
a relevant parameter in the black hole case. 
For a fixed value of $\ell/L$, we therefore compute the functional dependence of $M$ and $T$
upon $\bar{r}_{\rm h}$ numerically. Then, $S$ is also obtained by integrating the first
law of thermodynamics (\ref{first}).  

In the preceding subsection we observed that there is 
a critical point where the back reaction to the geometry becomes
important in the star case. The same is true for the black hole 
case. When the size of the central black hole is small, 
the temperature is high. Therefore the total mass is dominated 
by the radiation. As we increases the size of the black hole, 
radiation temperature drops. When the temperature drops down below  
$O(\ell^{-1/2} L^{-1/2})$, the effect of the radiation energy density 
becomes negligible in the same way as in the star case. 
When the size of the black hole is larger than that at the above
critical point, the geometry does not significantly deviate from 
the Schwarzschild AdS spacetime. 

As we further increase the size of the black hole, 
there appears another type of critical point,  
which does not exist in the star case. 
At this critical point, the stability of the system as a micro-canonical
ensemble changes. 
In this regime the total mass of the system can be approximated 
by the sum of the mass the black hole and that due to the CFT,
\begin{equation}
M\sim \bar{r}_{\rm h}+\frac{\pi^2}{30}g_{\rm eff}L^3 \bar{r}_{\rm h}^{-4}~, 
\end{equation}
where we used a rough estimate for the temperature, $T\sim 1/\bar{r}_{\rm h}$, which
is valid for $\bar{r}_{\rm h} \lesssim L$. 
The total mass takes a minimum at $\bar{r}_{\rm h}\sim  (\ell^2L^3)^{1/5}$. 
This means that there are several solutions with
the same total mass, but with different temperature or entropy. 
Amongst these solutions, the
one with the larger entropy
is micro-canonically stable. 
Since the total entropy is approximately given by 
\begin{equation}
S\sim \bar{r}_{\rm h}^2+\frac{\pi^2}{30}g_{\rm eff}L^3 \bar{r}_{\rm h}^{-3}~, 
\end{equation}
the sequence of the solutions is micro-canonically stable for $ \bar{r}_{\rm h} \gtrsim (\ell^2L^3)^{1/5}$.
The thermodynamic quantities at this point of the minimum mass can be estimated as
\bea
M/L&\sim& 
(\ell/L)^{2/5}~,\nonumber\\ 
\ell^{1/2}L^{1/2}T&\sim&(\ell/L)^{1/10}~,\nonumber\\
\ell^{-1/2}L^{-3/2}S&\sim&(\ell/L)^{3/10}~.\nonumber
\eea
Strictly speaking, we should 
consider a slightly different type of ensemble to discuss the
stability of the KR models. We discuss this issue later 
in Sec.~\ref{sec:AdS/CFT}.

\section{Numerical results}
\label{sec:numeric}

\begin{figure*}
\centering
\subfigure[CFT star]{\includegraphics[width=8cm, clip]{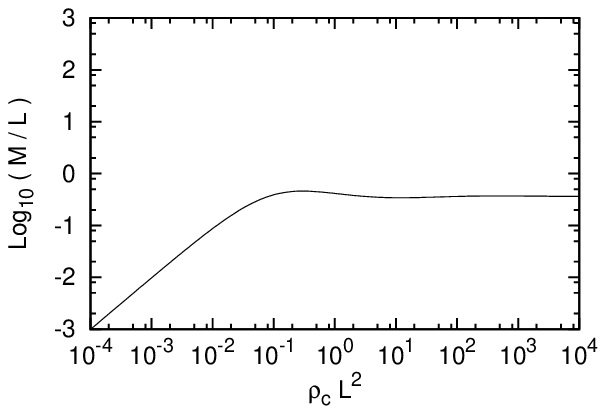}}
\subfigure[Black hole $+$ CFT]{\includegraphics[width=8cm, clip]{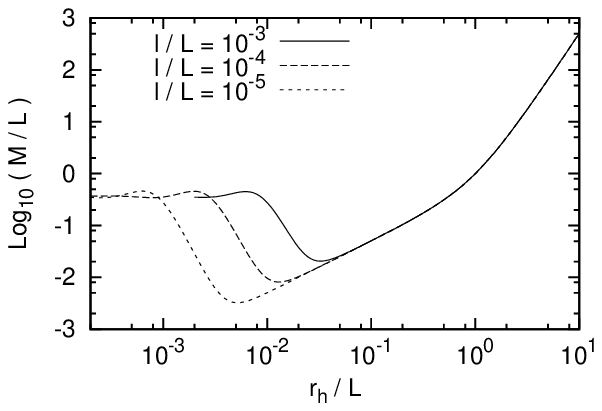}}
\caption{Total mass of CFT stars (left panel) and
 quantum black holes (right panel) with respect to the central
 density of the star and black hole horizon radius, respectively. In the
 right panel we set the parameter $\ell/L=10^{-3},10^{-4}$ and $10^{-5}$. The
 transition between the two sequences occurs at $M/L=0.36$.}
\label{M}
\vspace{4mm}
\end{figure*}

\begin{figure*}
\centering
\subfigure[CFT star]{\includegraphics[width=8cm, clip]{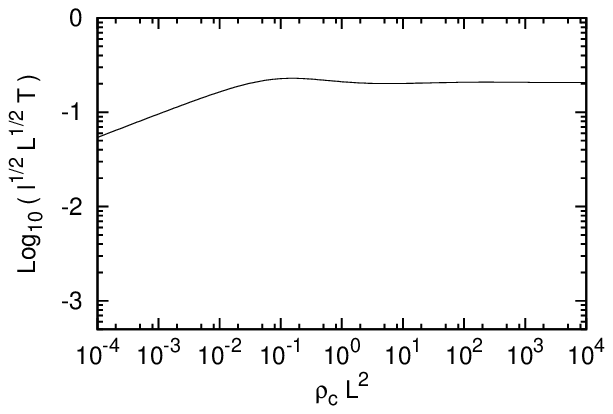}}
\subfigure[Black hole $+$ CFT]{\includegraphics[width=8cm, clip]{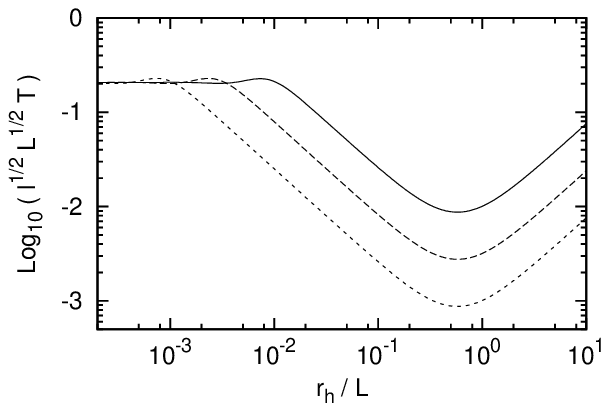}}
\caption{Plots for the temperature in the same way as in Fig.~\ref{M}. 
 The transition between the two sequences occurs at $\ell^{1/2}L^{1/2}T=0.21$.}
\label{T}
\vspace{4mm}

\end{figure*}

\begin{figure*}
\centering
\subfigure[CFT star]{\includegraphics[width=8cm, clip]{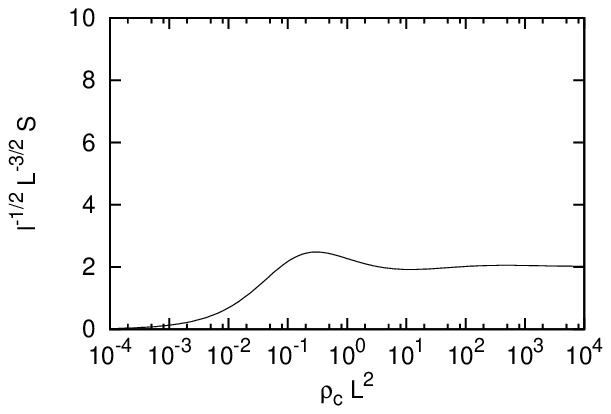}}
\subfigure[Black hole $+$ CFT]{\includegraphics[width=8cm, clip]{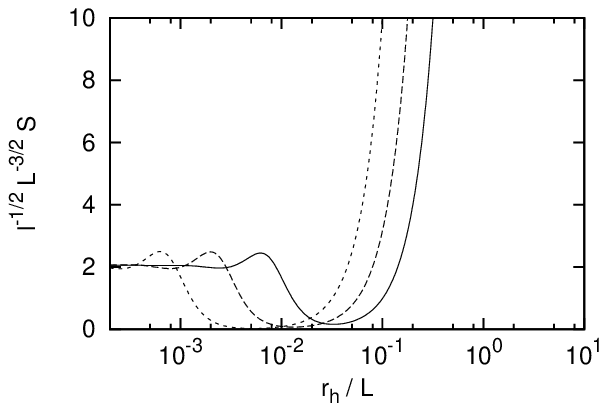}}
\caption{Plots for the temperature in the same way as in Fig.~\ref{M}. 
 The transition between the two sequences occurs at $\ell^{-1/2}L^{-3/2}S=2.0$.}\label{S}
\vspace{4mm}
\end{figure*}

In this section we will present the numerical results. We begin with
showing plots for the thermodynamic quantities: total mass $M$ (Fig.~\ref{M}), temperature $T$
(Fig.~\ref{T}) and entropy $S$ (Fig.~\ref{S}). The results for the CFT star and for the black holes are shown next to each other, illustrating
the smooth transition from one to the other. 

The two sequences are connected  
in the limit of infinite central density for the star configuration sequence, and in the limit of 
vanishing horizon radius for the black hole sequence.
A CFT star with large central density `becomes' a small mass black hole at the connection point. 
The transition occurs at 
\bea
M/L&=&0.36~,\nonumber\\
\ell^{1/2}L^{1/2}T&=&0.21~,\nonumber\\
\ell^{-1/2}L^{-3/2}S&=&2.0~.\nonumber
\eea 
These critical values do not depend on the ratio $\ell/L$.

As estimated in the preceding section, the minimum of the total mass for the quantum BH 
occurs at $\bar{r}_{\rm h} \sim (l^{2}L^3)^{1/5}$, which is consistent with the analytic estimate. 

\begin{figure*}
\centering
\subfigure[Phase diagram of quantum corrected black hole and CFT star]{\includegraphics[width=7.5cm, clip]{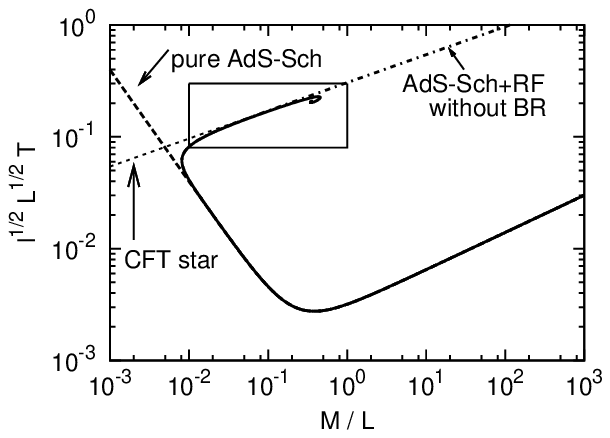}}
\subfigure[Closeup of the transition point]{\includegraphics[width=7.5cm, clip]{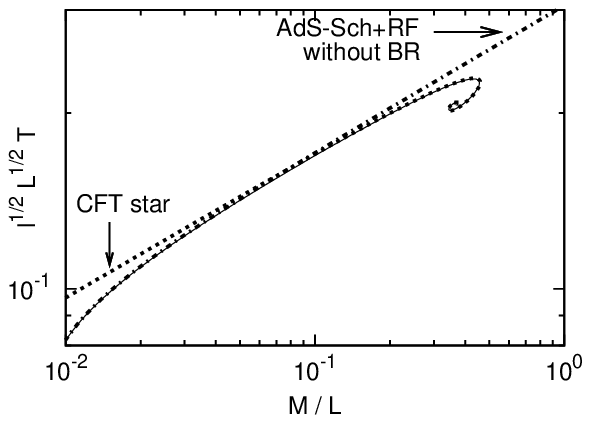}}
\caption{Relation between $M$ and $T$ for CFT stars 
 (dotted line) and quantum black holes (solid line). 
For the black hole system, we set
 $\ell/L=10^{-4}$. To understand the back reaction effect more clearly
 we add temperature-energy relation for Schwarzschild AdS space 
with (dotted-dashed line) and without (dashed line) 
the contribution of the radiation fluids. 
The right panel shows the closeup around 
 the transition point.}\label{PD}
\vspace{4mm}
\end{figure*}

Figure~\ref{PD} shows the relation between $M$ and $T$. 
The dotted line refers to the star sequence, 
while the solid line to the quantum black hole sequence. 
In order to clarify the back reaction effects, 
two additional reference curves are also shown in the same figure. 
The dashed line refers to the purely Schwarzschild AdS black hole case, 
and the dotted-dashed line refers to the sum of the black hole mass 
and the energy due to the CFT without taking into account the back 
reaction to the geometry. 
Figure~\ref{PD}, once again, shows the
smooth transition between the sequences of CFT stars and quantum black holes. 
The solid line starts to deviate from dotted-dashed
line at $\bar{r}_{\rm h}\sim \sqrt{\ell L}$, where the back reaction effects begin to
work. As is expected, the solid line deviates from pure Schwarzschild AdS case (dashed
line) when the energy of CFT becomes relevant at $M \sim
(\ell^2 L^3)^{1/5}$, corresponding to
$\bar{r}_{\rm h} \sim (\ell^2 L^3)^{1/5}$.

\begin{figure*}
\centering
\subfigure[CFT star]{\includegraphics[width=7.5cm, clip]{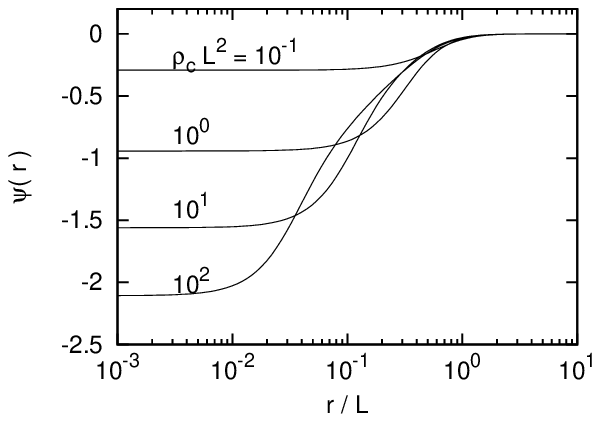}}
\subfigure[Black hole $+$ CFT]{\includegraphics[width=7.5cm, clip]{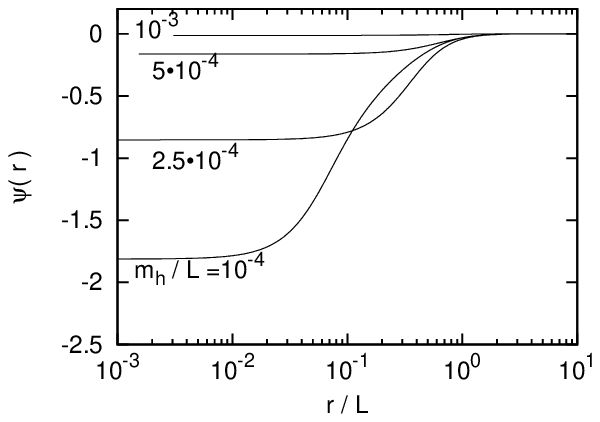}}
\caption{The metric function $\psi(r)$ for CFT stars (left
 panel) and for quantum black holes
 (right panel) for various central densities and black hole masses,
 respectively. We set $\ell/L=10^{-5}$.}
\label{psi}
\vspace{4mm}
\end{figure*}

\begin{figure*}
\centering
\subfigure[CFT star]{\includegraphics[width=7.5cm, clip]{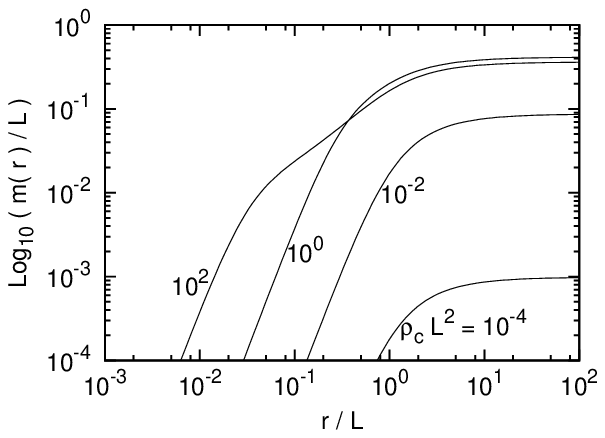}}
\subfigure[Black hole $+$ CFT]{\includegraphics[width=7.5cm, clip]{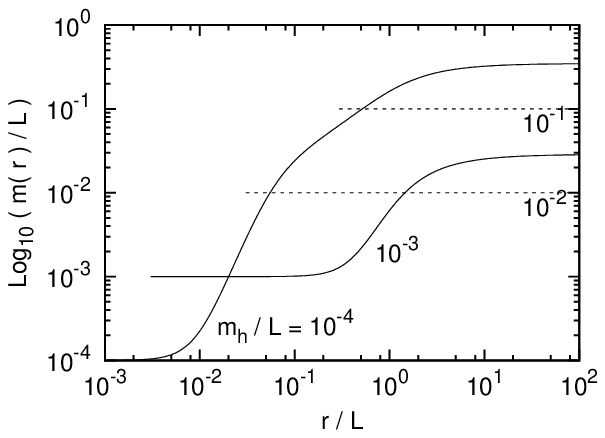}}
\caption{Plot for the metric function $m(r)$ in the same way as in 
Fig.~\ref{psi}.} 
\label{mass} 
\vspace{4mm}
\end{figure*}

In addition to the thermodynamic functions, we are interested in how the
CFT back reaction alters the spacetime geometry.  
Figures \ref{psi} and \ref{mass} show the behavior of the metric
functions $m(r)$ and $\psi(r)$ for CFT stars and quantum black holes. We
calculated them for various central densities $10^{-4} \leq \rho_c L^2
\leq 10^{2}$ in the star case and for various black hole masses 
$10^{-4} \leq m_{\rm h}/L\leq 1$ in the black hole case for $l/L=10^{-5}$. 
The results of the numerical computation are shown only for $r\geq r_{\rm in}$ for the black
hole sequence.

It is easy to see that the metric functions are almost the same 
for the star sequence in the large central density limit 
and for  the black hole sequence in the small size limit. 
Let us focus on, for example, the curves for $\rho_{\rm c} L^2=10^{2}$ in the left panel and $m_{\rm h}/L=10^{-4}$ in the right panel. First, for a small radius ($r/L \lesssim 10^{-2}$ for $\rho_{\rm c} L^2=10^{2}$ and $m_{\rm h}/L=10^{-4}$), there is a constant density core
described by 
\begin{equation}
\psi(r) \sim \text{const}~,
\end{equation}
and
\begin{equation}
\frac{m(r)}{L} \sim \text{const} \times \left( \frac{r}{L} \right)^3 .
\end{equation}
Here the constants are determined by the central density or the black hole
mass. 
This region is followed by the 
intermediate region ($10^{-2} \lesssim r/L \lesssim 1$). 
The behavior in this region is approximately obtained by 
solving Eqs.~(\ref{Eins1}) and (\ref{Eins3}) for $r/L\ll 1$, 
assuming power law solutions for $m$ and $\rho$. 
With the aid of Eqs.~(\ref{eos}), (\ref{Eins1}) and (\ref{Einspsi}), we obtain 
\begin{equation}
\frac{m(r)}{L} \sim \frac{3}{14}\left( \frac{r}{L} \right),
\end{equation}
and
\begin{equation}
\psi(r) \sim \frac{1}{2}\ln \left( \frac{r}{L} \right)~. 
\end{equation}

For stars with large central density and for black holes with small
mass, $\psi(r)$ takes a large negative value for a small $r$, 
which means a large red shift factor. 
The growth of red shift factor compensates the usual growth of black hole
temperature in the small black hole limit, and explains the convergence
of global temperature of the system.
We also mention that in the black hole case the back reaction effects are small
for $\bar{r}_{\rm h} \gtrsim (\ell^2L^3)^{1/5}$. In this case $m(r)$ takes
an almost constant (not small) value all over the spacetime (dotted lines in the
right panel of Fig.~\ref{mass}).

\section{AdS/CFT interpretation}
\label{sec:AdS/CFT}

In this section we will discuss the implications of our calculations for
black hole solutions in the KR model based on the AdS/CFT 
correspondence~\cite{Gubser2,Hawking,tanaka}. 
When we discuss the AdS/CFT correspondence in the KR model, 
our asymptotically AdS brane does not throughly surround the five-dimensional bulk space. Therefore, in addition to the CFT considered so far (CFT1), 
we need to include the contributions from another CFT residing on the boundary
that limits the other side of the bulk 
(CFT2)~\cite{KR2}. 
As long as thermal equilibrium state is concerned, we 
have to relate the temperature of CFT2 to that of CFT1. 
For convenience, we introduce a second brane at a finite but large
distance from the first brane. 
We calculate first the entropy of CFT2 on this second brane and then
send the brane separation to infinity.
The line element of four-dimensional metric induced on the second brane is 
approximated by the pure AdS metric:
\begin{equation}
ds^2=-f(r_2)dt_2^2+f(r_2)^{-1}dr_2^2+r_2^2d\Omega^2~,
\end{equation}
where $t_2$ and $r_2$, respectively, are time and radial coordinates on
the second brane, $f(r_2)\equiv 
1+r_2^2/L_2^2$, and $L_2$ is the AdS curvature length on the second
brane. 
The infinitesimal proper time interval for a static observer in these coordinates is
$f(r_2)^{1/2}dt_2$. On the 
other hand, this brane can be embedded in the five-dimensional
bulk, which behaves as (\ref{KRcoord}) in the asymptotic region. 
In the coordinates of Eq.~(\ref{KRcoord}) the second brane is located 
at $y=y_2$ approximately. 
The proper time interval in these coordinates is described by
$\ell\cosh (y_2/\ell)\,(1+\bar r^2)^{1/2}d\bar t$. 
Thus, the ratio between these two time coordinates is 
\begin{equation}
dt_2/d\bar t=dr_2/d\bar r=\ell\cosh(y_2/\ell)=L_2~.
\label{dt_2/dt}
\end{equation}
As for the first brane, a parallel discussion applies 
as long as a large radius limit is concerned. 
The induced metric on the first brane is also asymptotically AdS, 
and the location of the brane is also specified by a $y$-constant
surface there. 
Thus, we find $dt_2/dt=L_2/L$. 
Therefore, when thermal equilibrium is realized in the five-dimensional picture, 
the relation between the temperatures of CFT1 and CFT2 is given by
\begin{equation} 
TL=T_2L_2~.
\end{equation}
By using the radiation fluid approximation, the entropy of CFT2 can be estimated as 
\begin{eqnarray}
S^{CFT2}&=&\int^{\infty}_{0}4\pi r^2 \sqrt{g_{rr}}\, s\, dr
         ={\pi^5\over 2G_4} T^3L^3l^2,
\label{CFT2}
\end{eqnarray}
where $s=(4/3) (\pi^2/30)g_{\rm eff} (T_2/\alpha)^3$ is 
the radiation fluid entropy density with 
$g_{\rm eff}$ given in Eq.~(\ref{geff}). 
Since the entropy estimated in Eq.~(\ref{CFT2}) is independent of the
position of the second brane where the CFT2 lives, we 
send the second brane to the bulk boundary by taking the
limit $y_2\to\infty$. 

Now, let us consider brane-localized black holes which 
should correspond to four-dimensional asymptotically AdS 
quantum black holes. The AdS/CFT correspondence indicates that the
micro-canonical stability in the four-dimensional CFT picture should
correspond to the dynamical stability in the five-dimensional picture. 
(Notice that there is no reservoir of energy in the present system.) 
Hence, the previous discussion should be slightly modified by taking into
account the contribution from CFT2. 
Estimating the mass for the CFT2 taking into account a redshift factor, 
the total mass of the system will be given by 
\begin{eqnarray}
 M_{\rm tot} = M^{CFT1}\Bigl(\bar{r}_{\rm h},\ell,L
\Bigr)
 + {3\pi^5\over 8G_4} T^4\Bigl(\bar{r}_{\rm h},\ell,L \Bigr) L^3
		    \ell^2, \cr
\label{Mtot}
\end{eqnarray}
where the first term represents what we evaluated numerically in the
preceding section and the second term is the contribution from CFT2. 
A micro-canonical stable-unstable transition takes place 
at $\bar{r}_{\rm h} = 0.38\cdot (\ell^{2}L^{3})^{1/5}$, where $M_{\rm tot}$ above is 
minimized. 
Therefore the brane-localized black holes 
are expected to be stable (unstable) when the circumferential
radius of the brane cross-section of the horizon is larger (smaller)
than the above critical value.  

We can also discuss the possible shape of the corresponding five-dimensional solution. 
In the KR model, there is a black string solution, and its brane induced
geometry is exactly Schwarzschild AdS.  
Although this solution does not
satisfy the boundary condition that the metric should get close to the 
five-dimensional pure AdS at $y\to \infty$,  
we expect that, when the induced geometry of a brane-localized black hole 
is close to Schwarzschild AdS, the bulk geometry is
also close to a black string solution. 
As we showed in the preceding section, this is the case 
when the four-dimensional horizon radius is larger than 
$(\ell^{2}L^{3})^{1/5}$. In this case the metric function $m(r)$ is almost constant 
and $\psi(r)\sim 0$ for any $r$. 
(see Figs.~\ref{psi} and \ref{mass}). 
The AdS/CFT correspondence suggests that large
black holes should look like black strings in the five-dimensional
picture, but there is small deviation from the Schwarzschild AdS.
We expect that this small deviation is due to the truncation 
of the horizon of the ``black string'' at a finite distance far from the brane, 
having a cap there. Roughly speaking, the cap will be formed near the 
throat corresponding to $y=0$. 
In contrast, when $\bar{r}_{\rm h}\lesssim (\ell^{2}L^{3})^{1/5}$, the behavior of 
$m(r)$ and $\psi(r)$ is clearly different from the Schwarzschild AdS
case. 
This indicates that the five-dimensional bulk black hole dual to a
four-dimensional unstable black hole is not like a black string.

We can also estimate the expected size of the black hole in the five-dimensional
picture since 
the entropy is related to 
the five-dimensional area of black hole horizon $A_5$ by
\begin{equation}
S=\frac{A_5}{4 G_5}~. 
\end{equation}
We may define the corresponding five-dimensional horizon radius by  
\begin{equation}
r_{\rm h}\equiv \left({A_5\over 2\cdot2\pi^2}\right)^{1/3}~, 
\end{equation} 
where the factor $2$ represents the presence of two
floating black holes (one for each side of the bulk interrupted by the brane). 
The total entropy and hence the size of bulk floating black holes are 
almost constant in the course of the transition (see Fig.~\ref{S}). 
For example, 
the horizon radius of the five-dimensional black hole, $r_{\rm h}$, 
at the stability changing point is estimated as 
\begin{equation}
 r_{\rm h}= 0.7\cdot (\ell^3 L^2)^{1/5}, 
\end{equation}
from $\ell^{-4/5}L^{-6/5} G_4 S_{\rm tot}=3.4$ at $\bar{r}_{\rm h}=0.38\cdot(l^2L^3)^{1/5}$, where
\begin{eqnarray}
 S_{\rm tot} = S^{CFT1}\Bigl(\bar{r}_{\rm h},\ell,L
\Bigr)
  + {\pi^5\over 2G_4} T^3\Bigl(\bar{r}_{\rm h},\ell,L \Bigr) L^3
		    \ell^2. \cr
\label{Stot}
\end{eqnarray}
Again, the first term represents what we evaluated numerically in the
preceding section and the second term is the contribution from CFT2.

Let us now move on to 
the transition between the sequences of
floating black holes and brane-localized black holes. 
Corresponding to this transition, 
in the four-dimensional picture, we have confirmed that there is a
transition between CFT stars and 
quantum black holes. 
According to the results of our calculation, the transition occurs at 
\begin{eqnarray}
 \ell^{1/2} L^{1/2}T&=&0.21,\nonumber \\
 \ell^{-1/2}L^{-3/2}G_4 S_{\rm tot} &=& 2.0+1.4=3.4.
\end{eqnarray} 
Here, $2.0$ comes from CFT1 and $1.4$ from CFT2.  
These critical values are independent of the ratio $\ell/L$. 
From the entropy, the horizon radius of the five-dimensional black hole 
just touching the brane is estimated as 
\begin{equation}
 r_{\rm h}=0.7 \cdot (lL)^{1/2}.  
\end{equation}

\begin{figure}[h]
    \begin{center}
      \resizebox{80mm}{!}{\rotatebox{-90}{\includegraphics{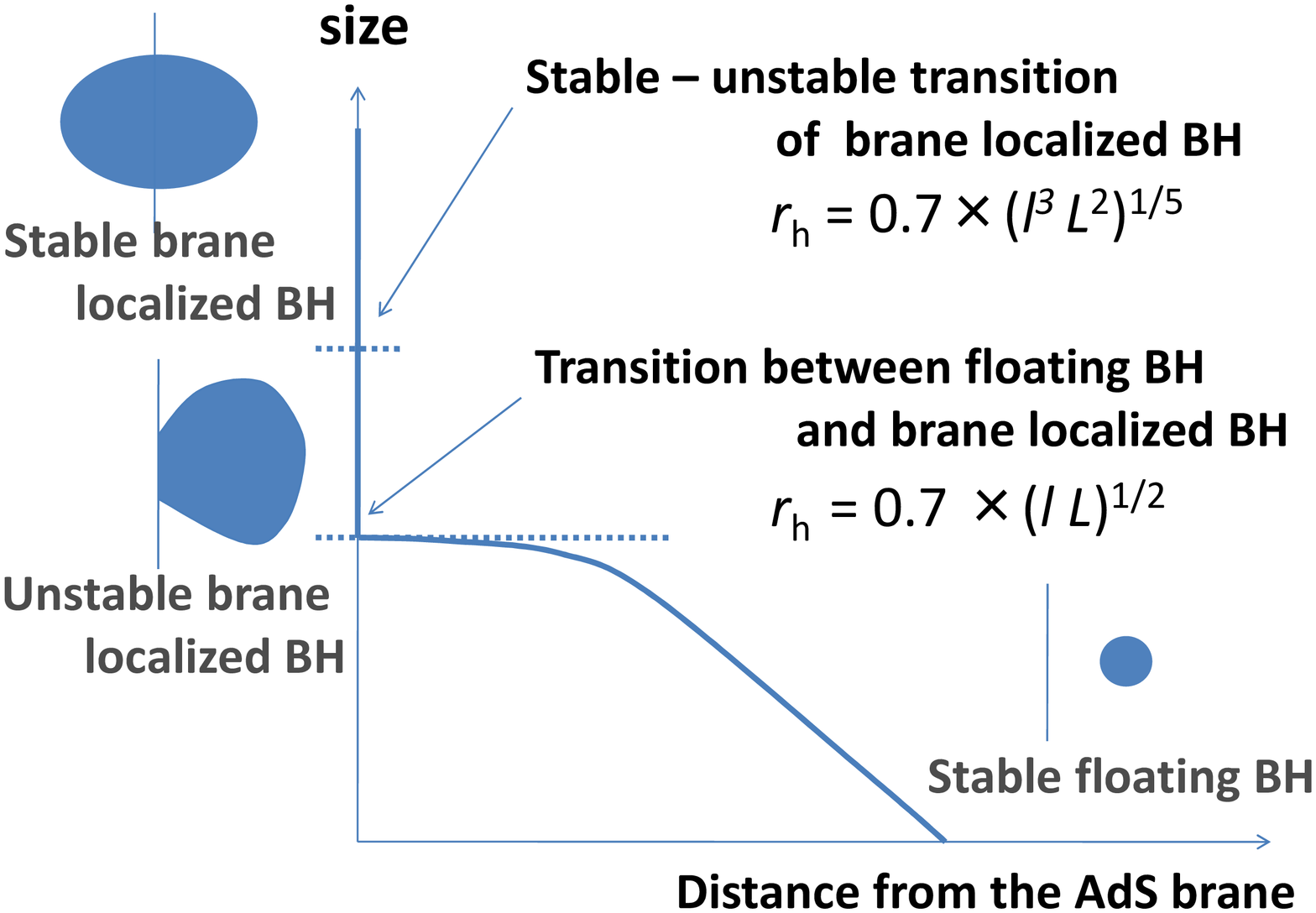}}} \\
  \caption{Phase diagram of BH solutions in the KR model.}\label{KRBH}
    \end{center}
\vspace{4mm}
\end{figure}

As we can see from Figs.~\ref{psi} and \ref{mass}, the geometry of a
star configuration in the large 
central density limit is very similar to that of a small black hole. 
This indicates that 
the five-dimensional geometry is also similar between the bulk floating black
holes just before touching the 
brane and the brane-localized black holes just after touching. The expected
phase diagram of black hole solutions in the KR model is illustrated in
Fig.~\ref{KRBH}.

Before closing this section, we would like to mention the screening
effect. As we have seen, if we use the Page's approximation instead 
of our more crude radiation fluid approximation, the mass varies logarithmically
in $r$ at infinity. In five-dimensional picture this phenomena 
can be understood as the leakage of gravitons from the brane on the 
CFT1 side because massless gravitons in the four-dimensional sense, 
which mediate the mass information to the infinity,  
are localized on the CFT2 side. Then, at a large distance
the leaked energy should be observed as the energy on the CFT2 side. 
In the four-dimensional CFT language this transmutation of energy 
from CFT1 to CFT2 can be correctly described only when 
the interaction between CFT1 and CFT2 is treated appropriately. 
However, in the fluid approximation we treated CFT1 and CFT2 
as completely independent components except for tuning the 
temperature. In such a treatment the energy transfer from CFT1 
to CFT2 is not taken into account. Therefore, when we identify the mass, 
we do not have to worry about 
the screening effect in this approximation.

\section{Summary}
\label{sec:summary}

We analyzed asymptotically AdS configurations with and without event
horizon in thermal equilibrium including the 
quantum back reaction due to CFT by using radiation fluid approximation 
with the aim to clarify the phase diagram structure of black objects in the KR model. 
We referred to the configurations with and without a horizon as 
CFT stars and quantum black holes, respectively. 
We have confirmed that the radiation fluid approximation is good when
typical length scales like the horizon radius  $\bar{r}_{\rm h}$ of the 
black hole are all larger than the bulk curvature scale $\ell$, 
in which the AdS/CFT correspondence is expected to be valid.
We calculated the metric and the thermodynamic quantities 
and found that: (i) the sequence of solutions of 
CFT stars is smoothly connected to the sequence of 
quantum black holes in the limit of infinite central density, (ii) the thermodynamically 
stable-unstable transition in the sequence of quantum black holes occurs
when the horizon radius $\bar{r}_{\rm h}$ is 
about $(\ell^2L^3)^{1/5}$, 
(iii) because of the back reaction effects, the temperature of the system 
converges to $\approx 0.21\cdot (\ell L)^{-1/2}$ in the limit $\bar{r}_{\rm h} \rightarrow 0$, 
(iv) for $\bar{r}_{\rm h} \gtrsim (\ell^2L^3)^{1/5}$, back reaction effects are
negligible and the space-time is approximately given by Schwarzschild AdS.

We also discussed the implications of our calculations for black hole solutions 
in the KR model based on the AdS/CFT correspondence. 
We claimed that (i) there are stability changing points along the 
sequence of brane-localized black hole solutions.  
The first transition corresponding to the minimum total mass of the system 
occurs when the five-dimensional horizon radius is 
$\approx 0.7\cdot(\ell^3L^2)^{1/5}$; (ii) the sequence of bulk floating black holes
leads to the sequence of brane-localized black holes and this transition
between these two sequences occurs when the black hole temperature is 
$\approx 0.21 \cdot (\ell L)^{-1/2}$ and the five-dimensional black hole
horizon radius is $\approx 0.7\cdot(lL)^{1/2}$.

\section{Acknowledgements}
This work is supported by the JSPS through Grants Nos. 19540285, 19GS0219, 2056381, 20740133, 21244033. 
We also acknowledge the support of the Global COE Program ``The Next
Generation of Physics, Spun from Universality and Emergence'' from the
Ministry of Education, Culture, Sports, Science and Technology (MEXT) of
Japan is kindly acknowledged.


\begin{thebibliography}{99}
\bibitem{Mal}
J.~M.~Maldacena, Adv.\ Theor.\ Math.\ Phys.\ {\bf 2} 231 (1998).
\bibitem{Wit}
E.~Witten, Adv.\ Theor.\ Math.\ Phys.\ {\bf 2} 253 (1998).
\bibitem{Gub}
S.~S.~Gubser, I.~R.~Klebanov, A.~M.~Polyakov, Phys.\ Lett.\ B \textbf{428} 105114 (1998).
\bibitem{RS}
L.~Randall, R.~Sundrum, Phys.\ Rev.\ Lett.\ \textbf{83} 4690 (1999).
\bibitem{GT}
J.~Garriga, T.~Tanaka, Phys.\ Rev.\ Lett.\ \textbf{84}, 2778 (2008).
\bibitem{Gubser2}
S.~S.~Gubser, Phys.\ Rev.\  D {\bf 63}, 084017 (2001) [arXiv:hep-th/9912001].
\bibitem{Hawking}
S.~W.~Hawking, T.~Hertog, H.~S.~Reall, Phys.\ Rev.\ D \textbf{62} 043501 (2000).
\bibitem{TanakaFRWGW}
T,~Tanaka,arXiv:gr-qc/0402068
\bibitem{PujolasDW}
L.~Grisa, O.~Pujolas, J.\ High Energy Phys.\ \textbf{0806} 059 (2008)
\bibitem{Duff}
M.~J.~Duff, J.~T.~Liu,  Phys.\ Rev.\ Lett.\ \textbf{85} 2052 (2000).
\bibitem{tanaka}
T.~Tanaka, Prog.\ Theor.\ Phys.\ Suppl.\ \textbf{148} 307 (2003).
\bibitem{Shiromizu}
T.~Shiromizu, D.~Ida, Phys.\ Rev.\ D \textbf{64} 044015 (2001).
\bibitem{GarrigaSasaki}
J.~Garriga, M.~Sasaki, Phys.\ Rev.\ D \textbf{62} 043523 (2000).
\bibitem{Kudo1}
H.~Kudoh, T.~Tanaka, N.~Nakamura, Phys.\ Rev.\ D \textbf{68} 024035 (2003).
\bibitem{Yoshino}
H.~Yoshino, J.\ High Energy Phys.\ {\bf 01} 068 (2009).
\bibitem{emparan}
R.~Emparan, A.~Fabbri, N.~Kaloper, J.\ High Energy Phys.\ \textbf{0208} 043 (2002).
\bibitem{KR}
A.~Karch, L.~Randall, J.\ High Energy Phys.\ \textbf{05} 008 (2001).
\bibitem{tanaka2}
T.~Tanaka, Prog.\ Theor.\ Phys.\ Suppl.\ \textbf{121} 1133 (2009).
\bibitem{flachitanaka}
A.~Flachi, T.~Tanaka, Phys.\ Rev.\ D {\bf 78}, 064011 (2008).
\bibitem{page}
D.~N.~Page, Phys.\ Rev.\ D \textbf{25} 1499-1509 (1982).
\bibitem{KR2}
A.~Karch, L.~Randall, Phys.\ Rev.\ Lett.\ \textbf{87} 061601 (2001).
\bibitem{gregory}
R.~Gregory, S.~F.~Ross, R.~Zegers, J.\ High Energy Phys.\ \textbf{09}, 029 (2008).
\bibitem{anderson}
P.~R.~Anderson, W.~A.~Hiscock, D.~A.~Samuel, Phys.\ Rev.\ D \textbf{51} 4337 (1995).
\bibitem{porrati}
M.~Porrati, J.\ High Energy Phys.\ \textbf{0204}, 058 (2002).
\bibitem{AdSDuff}
  M.~J.~Duff, J.~T.~Liu and H.~Sati, Phys.\ Rev.\  D \textbf{69} 085012 (2004).
\bibitem{3/4factor}
S.~S.~Gubser, I.~R.~Klebanov, A.~W.~Peet, Phys.\ Rev.\ D \textbf{54} 3915 (1996).
\bibitem{PP}
D.~N.~Page, and K.~C.~Phillips,~ Gen.~Rel.~Grav. \textbf{17} 1029 (1985).

\end{thebibliography}
\end{document}